# ORIGINAL PAPER – TITLE PAGE

## INDIRECT COMPARISONS FOR HEALTH TECHNOLOGY ASSESSMENT: A PRACTICAL METHODOLOGICAL GUIDE AND TIPS WITH INSIGHTS FROM THE FRENCH TRANSPARENCY COMMISSION

**Running Title:** Practical Guide to Indirect Treatment Comparisons

**Author information**

Louise Baschet, Horiana, Bordeaux, France (corresponding author : louise.baschet@horiana.com)

Ana Jarne, Horiana, Bordeaux, France

Matthias Monnereau, [1]Horiana, Bordeaux, France, [2] Université Paris-Saclay, CESP, INSERM U1018 Oncostat, labeled Ligue Contre le Cancer, Villejuif, France

Clémence Fradet, AstraZeneca, Courbevoie, France

Axel Benoist, AstraZeneca, Courbevoie, France

**Abstract:**

Context: Indirect treatment comparisons (ITC) are essential when direct head-to-head evidence is unavailable. Their reliability depends on rigorous methodological choices and careful assessment of underlying assumptions. Appropriate methodological choices can help address challenges such as cross-country variations in treatment practices, ethical constraints, and evolving treatment landscapes during trial conduct. This opinion and perspective paper provides practical guidance to strengthen the quality, robustness and accuracy of ITCs in the context of health technology assessment (HTA) in France.
Methods: A panel of experts in ITCs and French market access environment developed the present strategic guidance, informed by previous work reviewing HTA methodological guidelines and complemented by a systematic review of Transparency Committee opinions from the French National Authority for Health (HAS).
Results: Key considerations include early anticipation of ITCs, justification of potential confounding factors, and rigorous assessment of similarity and transitivity in randomized trial-based comparisons. In network meta-analysis, the structure of the evidence network should be adapted to the specific decision context. Population-Adjusted Indirect Comparisons require careful reporting and interpretation of the effective sample size. When evidence relies on non-randomized clinical trials, comparisons between single-arm studies and external control arms may be appropriate under different scenarios, depending on the feasibility of conducting subsequent randomized studies.
Conclusions: Robust and reliable ITCs require methods consistent with the validity of their assumptions and the strength of the available evidence. This practical guidance supports the development of rigorous ITCs to inform decision-making in complex medical contexts where direct comparisons are not feasible.

## Context

Randomized controlled trials (RCTs) are the gold standard for assessing the comparative efficacy of medical interventions. However, direct comparisons in RCTs can be difficult due to the rapid and often simultaneous development of multiple innovative therapies, and the rise in studies involving small patient populations, linked to disease rarity or target treatments. In cases where an RCT is not feasible or acceptable,  an indirect treatment comparison (ITCs) may be used (1,2). This applies, for example, when treatment practices vary by country, there are ethical constraints limiting global trials, and/ or evolving treatment landscapes during trials. ITCs are also valuable in some cases to complement the results of RCTs, or RCTs are limited by time or ethical concerns (3–5). In addition, observational data from clinical databases and health registries have opened new opportunities for ITCs when direct comparisons are not feasible (6,7).

In this paper, we provide detailed guidance to increase the quality and robustness of ITCs that are being used for decision making by Health Technology Assessment (HTA) bodies, with a specific focus on the Transparency Commission of the French National Health Authority (HAS TC).

We aim to delineate key recommendations for effectively anticipating and conducting ITCs in alignment with guidelines, and to identify when the ITC can be considered for HTA decisions. We suggest additional inputs to existing guidelines to optimize the use of ITC in decision-making. Specifically, we aim to:

- discuss the situations where an ITC can be considered,
- increase the quality of ITC considered by the HTA bodies (focusing on the TC),
- report examples of ITC that have been considered in TC opinions.

## Methods

An expert panel was composed of clinicians, methodologists, biostatisticians, and TC specialists with expertise in ITC and the French market access environment to provide strategic guidance. The panel's discussions were informed by prior work that pragmatically reviewed various HTA guidelines addressing the evolving challenges and advancements in the field of ITCs, alongside a systematic review of all HAS TC opinions published between 2021 and 2023 involving indirect comparisons (8). The insights presented here reflect the synthesis of expert opinion and perspectives and the contextual interpretation of existing regulatory and methodological frameworks.

## Results

*Pre-considerations*

HTA consideration of an ITC often hinges on factors that extend beyond the characteristics of the comparison itself. First, the non-feasibility of a direct comparison using available data and context must always be assessed and justified. Second, HTA bodies examine whether the ITC was planned from the outset (especially in single-arm or placebo-controlled trials), or whether it was considered retrospectively. ITCs based on RCT enabling anchored comparisons must be distinguished from those that are not, as this affects design, potential biases and reliability of evidence. Prior discussions with HTA bodies to present and justify the ITC approach are useful when possible.

*Comparisons based on randomized trials (anchored ITCs)*

There are three general principles for ITCs that can be used to complement the results of RCTs for regulatory bodies:

1. **Anticipate the ITC**: This is explicitly requested by HAS. An ITC may become necessary, like concurrent drug development or evolving standards of care. When possible, consider defining specific endpoints or even subpopulations aligned with expected future analyses.

2. **Justify confounding factors**: Anchored ITCs primarily rely on the identification of treatment effect modifiers (TEM). A thorough literature review, or at least, analysis of TEM from RCT subgroup data, is essential. Expert opinion should be supported by relevant evidence, including grey literature and trial subgroup analyses.

3. **Clarify exchangeability assumptions**: Challenges in defining exchangeability (combining heterogeneity, similarity and transitivity) are common (9). In the pragmatic review of TC opinions (8), lack of homogeneity was the most frequent TC concern across all ITCs, though distinctions between these concepts are often unclear in TC decisions. These properties are interrelated (10): failures in homogeneity or consistency (variation within or between trials) often stem from imbalances in TEM between studies (i.e., a violation of similarity). Heterogeneity is frequently mentioned in recommendations, yet thresholds for ITC feasibility, model choice (e.g. fixed vs. random effects), and conditions for using informative priors (which may help address convergence issues or data limitations) remain undefined. We recommend explicitly defining and distinguishing these concepts in HTA submissions, and systematically justifying methodological choice..

Four specific situations within RCTs also should be considered:

1. **Comparisons based on network meta-analysis (NMA)**: Adapt the network to the context. Globally developed pharmaceutical companies often use comprehensive networks including many international treatments. However, when presenting to HTA bodies across different countries, some treatments may be irrelevant locally. The NMA can be limited to relevant comparators, though opinions diverge. Our experience supports constructing a minimal yet sufficient network.

2. **Anchored Population-Adjusted Indirect Comparisons (PAIC):** Presentation and interpretation of the Effective Sample Size (ESS) are critical. In feasibility studies, method choice should not solely rely on ESS. While the acceptable level of ESS loss is subjective, a substantial decrease signals underlying issues. In all cases, thorough testing and clear justification of underlying assumptions (such as population comparability) greatly improve acceptance and confidence in results.

3. **When an RCT with a comparator is not appropriate**: The most appropriate comparator depends on study timing and location. Inappropriate comparators were cited in 14.1% of TC opinions on RCT-based ITCs (8). To address this:

- Decisions should never favor placebo over an available active treatment solely to obtain market authorization.
- When other molecules (either in development or already on the market) are known that could serve as potential comparators, anticipating the collection of relevant outcomes or measures of interest from parallel studies can facilitate future comparisons.
- If a new comparator is published during the study and is considered to have a high potential to become the new standard of care, the relevance of the ongoing study may be compromised. In such cases, it is important to assess whether continuing and presenting the current comparison remains appropriate, or whether the study should be adapted through a protocol amendment to update the comparator, despite the potential impact on study design.
- If a new treatment targeting a specific population emerges during the study, a protocol amendment to include a comparable subpopulation may be considered.
- Planning larger studies than initially required can facilitate comparisons with emerging standards of care, dependent on cost and time constraints. To minimize the need for major redesigns, continuous monitoring is essential to anticipate changes in relevant comparators.

4. **Request for reevaluation**: Three cases may arise. First, when multiple treatments are deemed of equal value, with no clear preference, the objective is to show there are several viable options depending on individual and clinician/patient preferences. Second, if the index treatment has demonstrated superiority, the aim is to present this clearly. Finally, if a new treatment has surpassed the index treatment, it is still necessary to demonstrate its continued role in therapeutic strategy.

Figure 1 outlines a step-by-step process to select the most suitable ITC for RCT-based ITCs, thus maximizing quality.

**Comparisons based on non-randomized trials**

When an RCT is not feasible, the best compromise is to proceed with an externally controlled trial. This is possibly the most common scenario, and it is important to recall the TC position on this matter. (5) Focusing on a successful historical case, a strong indirect comparison was conducted in the TOSCA study (ClinicalTrials.gov ID : NCT05302297) for the reevaluation of Libtayo in 2023. This included a comprehensive consideration of confounding factors / effect modifiers, allowing for an ASMR (Improvement in Actual Benefit) IV (minor improvement) during reevaluation, instead of an ASMR V (no improvement) in the initial submission.

In other cases, consideration of Phase II single-arm studies, particularly for early access authorizations, depends on whether a confirmatory RCT is planned or ongoing. If such a study is underway, expectations focus on its results, and promising findings from an ITC based on the single-arm study may be acceptable while awaiting randomized evidence. If a confirmatory study has not yet started, concerns arise regardless of ITC methodology. Expectations in these cases are shaped by the contentious history of FDA accelerated approvals and subsequent policy shifts. (11) Moreover, if a confirmatory study is not feasible, this situation would not meet the FDA's proposal to accept supported data from separate sources as confirmatory evidence, even when results appear clearly positive.

Two recent and relevant documents regarding comparisons based on non-randomized trials have recently been published. Firstly, the Coordination Group on Health Technology Assessment (HTA CG) have proposed guidelines at the European level, (10,12) which incorporate the recommendations established by the European Network of Health Technology Assessment (EUnetHTA). (13) National guidelines were published in 2023 by a panel of experts commissioned by the French Ministry of Health, following consultations with patient associations, academics, manufacturers, and various institutions.(5) They emphasize that when ITC are required due to the absence of direct evidence from RCTs, the uncertainty regarding the estimation of treatment effects increases, although such methods may be acceptable under justified circumstances. The guidelines also provide a methodological framework, which may be difficult to implement in practice, particularly due to challenges in anticipating evidence needs. Beyond methodological considerations, the context in which the ITC is conducted is also important. As with ITC based on RCT, the choice of external comparator must reflect the most appropriate comparator.

The most acceptable ITC approaches are those that use Individual Patient Data. These can align with the FDA's guidelines for non-randomized studies(14) and the EMA's guidelines, (10,12). Unanchored MAICs are highly discouraged as it is very difficult to verify the assumption of conditional constancy of absolute effects. However, in 2021 they were found in the case of the requests for reevaluation for Kymriah (15) and Yescarta (16), both within the context of lymphomas, where the results obtained with this method were considered by the TC while highlighting the limitations. Key aspects, such as anticipation and the quality of the data used are improving. It is clear that gaining more experience in the field of comparative effectiveness studies will help advance drug development. There are still regular calls to raise and better control the standards for these observational studies.(11)

The use of observational or health system data to assess adverse events is complex and limited. Tolerability data relevant to clinical decision-making are scarce, often inconsistent with trial evidence, and biological adverse events are difficult to capture systematically. In TC opinions, ITCs mainly focus on efficacy, with tolerability rarely assessed using comparable rigor. No TC evaluation has yet used ITC for tolerability, though this may evolve with telemedicine and research on immune-related adverse events as surrogate markers.

As TC assessments rely on overall benefit/risk balance, analyses should support a favorable balance even without fully quantitative evaluation of adverse events. However, many ITCs overlook tolerability despite its importance. In some cases, such as high-risk treatments (e.g., anticoagulants), precise benefit/risk quantification is necessary. This requires ensuring comparability between groups for adverse events, accounting for confounders distinct from those affecting efficacy. While challenging, this may be feasible using observational data (e.g., hospitalization records for bleeding). In other contexts, a qualitative approach may be more appropriate given the multidimensional nature of tolerability.

Improving tolerability assessment in ITCs may involve greater use of quality-of-life measures, Patient-Reported Experience Measures (PREMs) collected through decentralized tools, or clinical biomarkers when reliable surrogate endpoints exist.

## Discussion and perspectives

ITCs present both challenges and opportunities for evaluating therapeutic benefit and positioning treatments within therapeutic strategies. Their underlying assumptions should guide method selection, ensuring alignment with the evolving landscape of evidence-based medicine.

For RCT-based comparisons, several steps can improve quality. NMAs are generally preferred when supported by robust feasibility assessments and clear justifications for exchangeability. When NMAs are not feasible, focusing on relevant sub-populations and considering PAICs can strengthen the credibility of results.

For non-randomized studies, external control arms often represent the most appropriate compromise, although the strength of the evidence depends not only on study design but also on contextual factors, particularly whether a confirmatory study is planned or ongoing.

Drawing primarily on HAS recommendations, this work aims to provide practical guidance for manufacturers navigating the methodological and strategic challenges of ITC. The feasibility and impact of these recommendations vary. Some, such as prospectively planning an external control arm or conducting an early systematic literature review (SLR) to identify appropriate comparators and relevant prognostic or effect-modifying factors, are simple to implement. Others, such as increasing sample size in anticipation of a future indirect comparison, are often difficult or impossible to implement in practice.

## Conclusion

This guidance serves as a practical resource for the preparation of submissions to HTA bodies, especially the French HAS TC. By following the outlined recommendations and adapting to the specific context of each evaluation, the quality of ITC can be enhanced, which can have a major impact on the ASMR rating and, consequently, on the treatment's positioning within the therapeutic strategy.

## Conflicts of interest

This study was supported by AstraZeneca.

## Figures

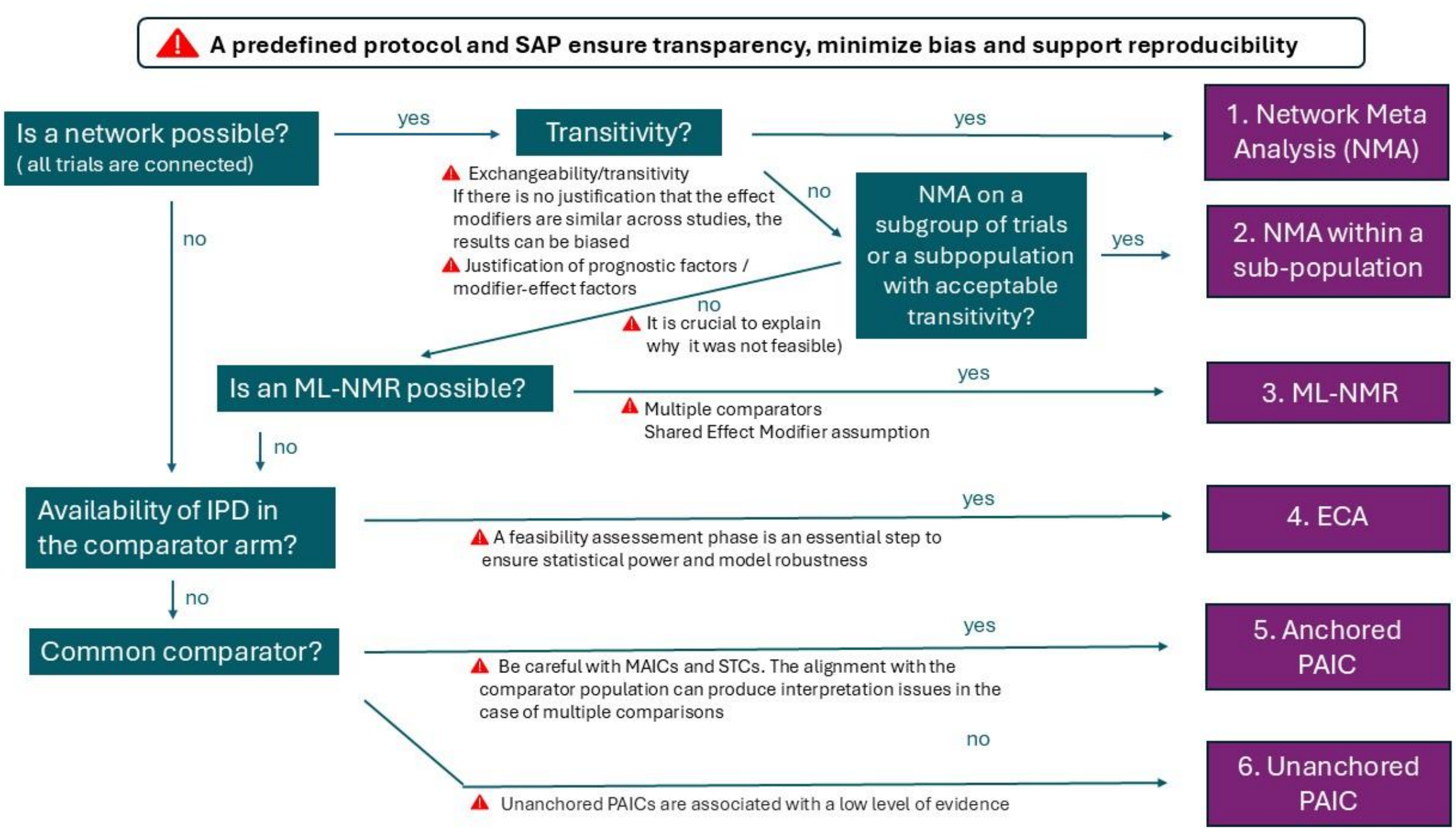

* Abbreviations: ECA: External Control Arm; IPD: Individual Patient Data; MAIC: Matching Adjusted Indirect Comparison; ML-NMR: Multi-Level Network Meta Regression; NMA: Network Meta-Analysis; PAIC: Population Adjusted Indirect Comparison; SAP: Statistical Analysis Plan; STC: Simulated Treatment Comparisons.

**Figure 1.** Step-by-step Process for Selecting the Most Appropriate Indirect Comparison